\documentclass[aps,prl,superscriptaddress,nofootinbib,twocolumn,amsmath,amssymb,longbibliography,floatfix]{revtex4-1}

\usepackage[T1]{fontenc}

\usepackage{graphicx}
\usepackage{dcolumn}
\usepackage{bm}
\usepackage[colorlinks]{hyperref}
\usepackage{amsmath}
\usepackage{makecell}
\usepackage[mathscr]{eucal}
\usepackage{multirow}
\usepackage{graphicx}
\usepackage{float}
\usepackage{subfigure}
\usepackage{tabularx}
\usepackage{booktabs}
\usepackage{colortbl}
\usepackage[table]{xcolor}
\usepackage{siunitx}

\definecolor{lightyellowgreen}{RGB}{216, 214, 194}
\definecolor{lightyellowgreen2}{RGB}{235, 234, 222}

\begin{document}


\title{Sample-half-inserted quantum interferometer}

\author{Wei Li}
\thanks{These authors contributed equally to this work.}
\affiliation{Key Laboratory of Low-Dimensional Quantum Structures and Quantum Control of Ministry of Education, Department of Physics and Synergetic Innovation Center for Quantum Effects and Applications, Hunan Normal University, Changsha 410081, China}
\affiliation{Hubei Key Laboratory of Optical Information and  Pattern Recognition, Wuhan Institute of Technology, Wuhan 430205, China}
\author{Tao Xie}
\thanks{These authors contributed equally to this work.}
\affiliation{Key Laboratory of Low-Dimensional Quantum Structures and Quantum Control of Ministry of Education, Department of Physics and Synergetic Innovation Center for Quantum Effects and Applications, Hunan Normal University, Changsha 410081, China}
\affiliation{Hubei Key Laboratory of Optical Information and  Pattern Recognition, Wuhan Institute of Technology, Wuhan 430205, China}
\author{Yu-Hang Luo}
\thanks{These authors contributed equally to this work.}
\affiliation{Hubei Key Laboratory of Optical Information and  Pattern Recognition, Wuhan Institute of Technology, Wuhan 430205, China}

\author{Kang Zheng}
\affiliation{Hubei Key Laboratory of Optical Information and  Pattern Recognition, Wuhan Institute of Technology, Wuhan 430205, China}
\author{Meiyu Peng}
\affiliation{Key Laboratory of Low-Dimensional Quantum Structures and Quantum Control of Ministry of Education, Department of Physics and Synergetic Innovation Center for Quantum Effects and Applications, Hunan Normal University, Changsha 410081, China}
\author{\mbox{Hui Yang}}
\affiliation{Key Laboratory of Low-Dimensional Quantum Structures and Quantum Control of Ministry of Education, Department of Physics and Synergetic Innovation Center for Quantum Effects and Applications, Hunan Normal University, Changsha 410081, China}
\author{Chunling Ding}
\affiliation{Hubei Key Laboratory of Optical Information and  Pattern Recognition, Wuhan Institute of Technology, Wuhan 430205, China}
\author{Chen-Zhi Yuan}
\affiliation{Hubei Key Laboratory of Optical Information and  Pattern Recognition, Wuhan Institute of Technology, Wuhan 430205, China}
\author{Omar S. Maga\~na-Loaiza}
\affiliation{Quantum Photonics Laboratory, Department of Physics and Astronomy, Louisiana State University, Baton Rouge, Louisiana 70803, USA}
\author{\mbox{Keyu Xia}}
\affiliation{College of Engineering and Applied Sciences and National Laboratory of Solid State Microstructures, Nanjing University, Nanjing 210023, China}
\affiliation{Shishan Laboratory, Suzhou Campus of Nanjing University, Suzhou 215000, China}
%
\author{Ryosuke Shimizu}
\affiliation{University of Electro-Communications, Chofu, Tokyo 182-8585, Japan}
%
\author{Hui Jing}
\email{jinghui@hunnu.edu.cn}
\affiliation{Key Laboratory of Low-Dimensional Quantum Structures and Quantum Control of Ministry of Education, Department of Physics and Synergetic Innovation Center for Quantum Effects and Applications, Hunan Normal University, Changsha 410081, China}
\affiliation{College of Science, National University of Defense Technology, Changsha 410073,China}

\author{Chenglong You}
\email{cyou2@lsu.edu}
 \affiliation{Quantum Photonics Laboratory, Department of Physics and Astronomy, Louisiana State University, Baton Rouge, Louisiana 70803, USA}
\author{Rui-Bo Jin}
\email{jrb@hunnu.edu.cn}
\affiliation{Key Laboratory of Low-Dimensional Quantum Structures and Quantum Control of Ministry of Education, Department of Physics and Synergetic Innovation Center for Quantum Effects and Applications, Hunan Normal University, Changsha 410081, China}
\affiliation{Hubei Key Laboratory of Optical Information and  Pattern Recognition, Wuhan Institute of Technology, Wuhan 430205, China}

\date{\today}
\begin{abstract}
Quantum technologies have been widely recognized as unprecedented opportunities for ultra-high precision metrology.  As a celebrated example in modern quantum optics, the Hong–Ou–Mandel (HOM) interferometer is well-known for enabling temporal resolutions on the attosecond scale.
However, the relatively low Fisher information per trial in ordinary HOM measurements typically necessitates tens of thousands of repetitions to achieve such precision. Here, we propose and demonstrate a sample-half-inserted HOM (SHOM) interferometer, which enhances the Fisher information by five orders of magnitude in a single interference event. By introducing an asymmetric photon-sample interaction, the SHOM configuration produces a distinctive dip–bump–dip interference structure, converting what was previously viewed an artifact into a helpful metrological resource. Experimentally, we measured the optical path  difference with an average precision of 4.09 nm (13.63 as) and an average accuracy of 1.22 nm (4.07 as)  using $O(10^7)$ photons.
Our results establish SHOM interferometry as an efficient phase-insensitive approach, not only paving the way toward practical quantum-enhanced thickness measurement for transparent materials, but also serving as an elegant strategy to improve the performance of various quantum devices.
\end{abstract}

\maketitle
Hong-Ou-Mandel (HOM) interferometry, a two-photon quantum interference effect, is widely regarded as a foundational technique in quantum optics \cite{Hong1987}. In a HOM interferometer, two identical photons are directed into the input ports of a beam splitter. Due to the bunching behavior of bosons, these two photons tend to exit together from the same output ports of the beam splitter. This distinctive quantum behavior results in the complete suppression of coincidence counts, manifesting as the characteristic HOM dip observed in coincidence measurements. Notably, the HOM effect extends beyond photons and has been observed across various quantum systems, including atoms \cite{Lopes2015Nature}, electrons \cite{Bocquillon2013Science}, plasmons \cite{Fakonas2014NP},  phonons\cite{Toyoda2015Nature}, and magnons \cite{Su2022PRL}, highlighting its broad applicability in the quantum regime. 

Undoubtedly, HOM interference has found extensive applications in quantum information technologies \cite{Jin2024PQE}. Over the past three decades, significant efforts have been devoted to harnessing HOM interference in quantum metrology, aiming to estimate physical quantities with sensitivities that surpass classical limits \cite{JordanPhysRevA2022}.  Owing to its intrinsic insensitivity to phase fluctuations and its natural dispersion cancellation properties, HOM interference has been widely adopted in various quantum sensing applications, including quantum optical coherence tomography (QOCT) \cite{Abouraddy2002PRA, Nasr2003PRL}, quantum ranging \cite{Triggiani2023PRAppl, Triggiani2024PRL}, thickness measurements \cite{Branning2000PRA, meskine2024approaching}, and quantum microscopy \cite{Ndagano2022NP}. A particularly notable demonstration by Branning et al. employed HOM interference to measure the thickness of birefringent crystals with a resolution of 0.1 fs (33 nm) \cite{Branning2000PRA}. While their common-path configuration offered excellent stability, it was restricted to birefringent samples. Subsequent developments led to the widespread adoption of HOM-based QOCT, which uses a two-arm interferometric configuration capable of measuring both birefringent and non-birefringent materials \cite{Nasr2003PRL, Nasr2008OE, Hayama2022OL,gracianoInterferenceEffectsQuantumoptical2019,dabrowskaQuantuminspiredOpticalCoherence2024,lavoieQuantumopticalCoherenceTomography2009,carrasco2004enhancing,ohOptimalGaussianMeasurements2019,kimDistributedQuantumSensing2024}. 
However, the achievable resolution in QOCT remains limited to the femtosecond regime.

\begin{figure*}[!ht]
\centering
\includegraphics[width=0.99\textwidth]{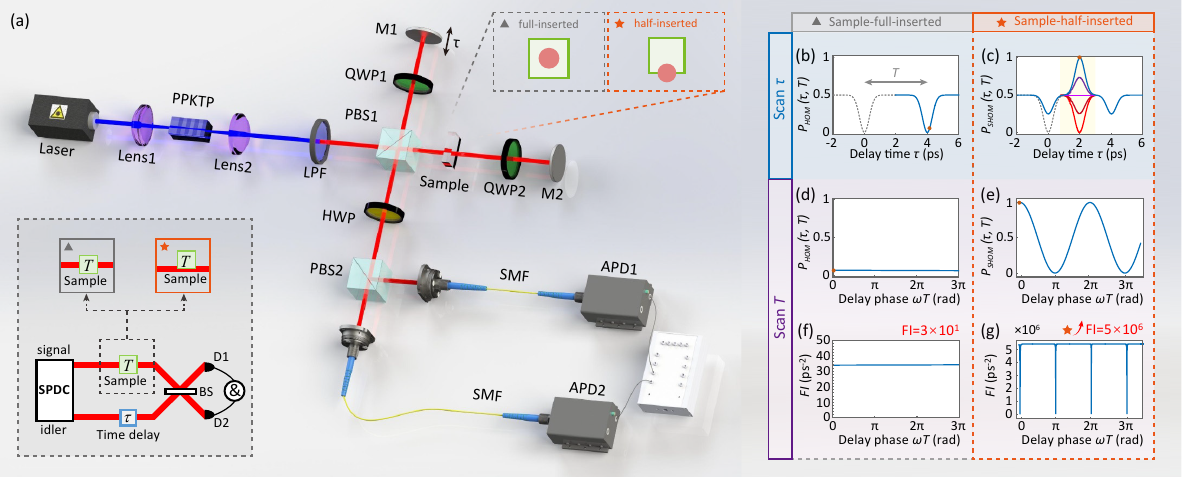}
\caption{\textbf{The concept of the sample-half-inserted HOM (SHOM) interference.} (a) Experimental setup for the SHOM interferometer. A narrowband laser with a center wavelength of 405.0(1) nm pumps a PPKTP crystal to produce photon pairs via type-II spontaneous parametric down-conversion (SPDC). The orthogonally polarized signal and idler photons are directed into a delay system consisting of a polarization beam splitter (PBS1), two quarter-wave plates (QWP1, QWP2) set at $45^\circ$, and two mirrors (M1, M2). Mirror M1 is mounted on a stepper motor to scan the optical delay $\tau$. After the delay line, the photons pass through a half-wave plate (HWP) set at $22.5^\circ$ and a second PBS (PBS2) before being coupled into single-mode fibers (SMFs) and detected by avalanche photodiodes (APDs) connected to a time interval analyzer (TIA). A glass sample can be inserted into the signal arm either fully or partially, as illustrated in the enlarged top-right and bottom-left inset. (b, c) Coincidence probabilities $P_{\rm HOM}(\tau, T)$ and $P_{\rm SHOM}(\tau, T)$ as functions of delay $\tau$, before (gray) and after (colored) inserting the sample with fixed sample-induced delay $T$. The different colors of the central structure in (c) correspond to different delay $T$. (d, e) Coincidence probabilities as functions of  delay $T$ with fixed delay $\tau$, sweeping $\omega T$ from 0 to 3$\pi$, where $\omega$ is the angular frequency of the biphotons. (f, g) Fisher information calculated from the interference patterns in (d) and (e). }
\label{Fig:concept}
\end{figure*}

In 2018, Lyons and co-workers demonstrated the first attosecond-resolution HOM interferometry \cite{Lyons2018SA}. This remarkable precision, however, came at the cost of repeating the measurement over 35,000 times using $O(10^{11})$ photons. 
As with other HOM-based schemes, a fundamental limitation remains: the relatively low Fisher information obtainable from a single interference event. Consequently, achieving high precision requires a large number of photons in repeated trials to accumulate sufficient statistical information. Here, we propose and experimentally demonstrate a novel strategy for significantly enhancing the Fisher information obtained from a single HOM interference measurement. Different from the conventional approach, where the sample is fully inserted into the optical path, we introduce a sample-half-inserted HOM (SHOM) interferometer. This modified configuration increases the Fisher information by up to five orders of magnitude relative to the standard HOM setup, leading to an improvement in measurement precision by approximately two orders of magnitude. In particular, we achieved an average precision of 4.09~nm (13.63~as) and an average accuracy of 1.22~nm (4.07~as) using only $O(10^7)$ photons.
Throughout this work, we define precision as the standard deviation of the measured values and accuracy as the difference between the estimated value and the known reference value.

\begin{figure*}[tbh]
\centering
\includegraphics[width=1\textwidth]{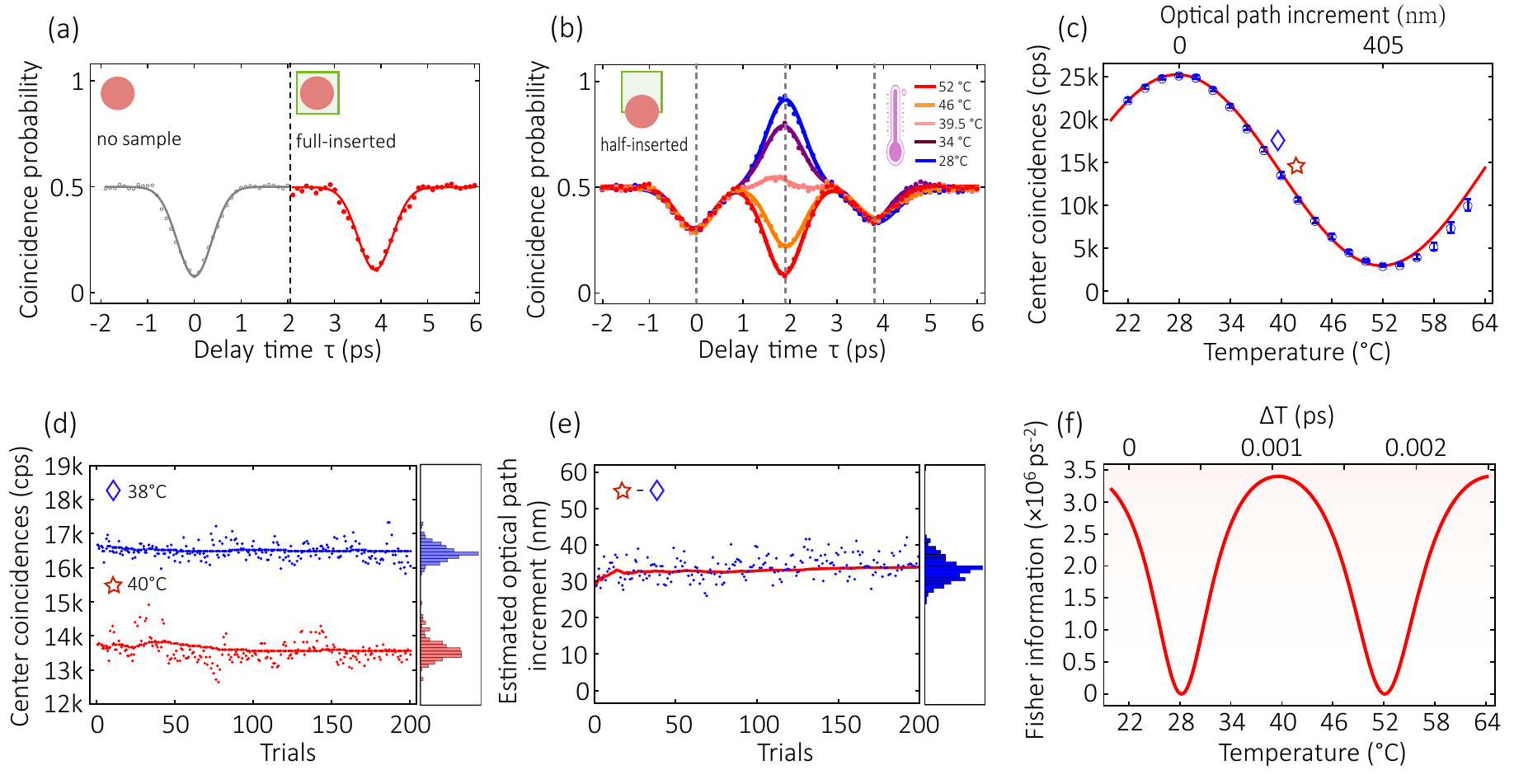}
\caption{\textbf{Experimental results of sample-half-inserted HOM interference.} (a) Hong–Ou–Mandel (HOM) interference patterns without the sample (gray) and with the sample fully inserted (blue). Dashed points represent experimental data, and solid lines denote Gaussian fits. (b) HOM interference patterns with the sample half inserted, where the sample is heated to different temperatures. (c) Coincidence counts at the interference center as a function of temperature, with time delay $\tau$ fixed at the central dip position from (b). (d) Coincidence counts at the interference center for \qty{38}{\degreeCelsius} (blue) and \qty{40}{\degreeCelsius} (red) measured over 200 trials. Solid lines represent the cumulative average; the histogram on the right illustrates the distribution of counts. (e) Estimated optical path increment obtained by subtracting the coincidence data at \qty{38}{\degreeCelsius} from that at \qty{40}{\degreeCelsius}, based on the results in (d). In (d, e), a trial corresponds to coincidence counts in 1 s.   (f) Calculated Fisher information based on the data in (c). 
}
\label{Fig:result}
\end{figure*}

The concept of our SHOM interferometer is illustrated in Fig. \ref{Fig:concept}. In a conventional HOM interferometer, as depicted in the bottom left of Fig. \ref{Fig:concept}(a), the signal and idler photons are directed to a 50:50 beam splitter. By scanning the delay time $\tau$ in the idler arm and recording the coincidence counts, a characteristic HOM dip is observed at $\tau=0$, as shown by the dashed gray line in Fig. \ref{Fig:concept}(b). When the sample is fully inserted into the signal arm, the HOM dip shifts from $\tau = 0$ to $\tau = T$, as shown by the solid blue line in Fig. \ref{Fig:concept}(b). Here, $T$ represents the additional propagation time introduced by the sample. As shown in the Supplementary Material (SM) \cite{note2} Section I \& II, the coincidence probability $P_{\rm HOM}(\tau ,T)$ as a function of the $\rm \tau$ and T can be expressed as 
\begin{equation}\label{Eq:HOM}
 P_{\text{HOM}}(\tau, T ) =   \frac{1}{2}   - \frac{1}{2}\exp[ - \frac{{\sigma _ - ^2{(\tau-T)^2}}}{2}].
\end{equation}
Here, $\sigma _ -$ is the spectral width of the biphoton in the $\omega_s -\omega_i$ direction, and $\omega_s$ and $\omega_i$ are the angular frequency of the signal and idler (see also SM Section II).

In contrast, when the sample is half-inserted into the signal arm, in which that only half of the signal photons pass through the sample while the other half propagate through air, the interference pattern changes significantly. Under this condition, the coincidence measurement exhibits two dips and a central structure, as illustrated in Fig. \ref{Fig:concept}(c). The coincidence probability $P_{\text{SHOM}}(\tau, T)$ as a function of the delay time $\tau$ and the sample-induced delay $T$ can be expressed as
\begin{equation}\label{Eq:SHOM}
\begin{split}
  &P_{\text{SHOM}}(\tau, T ) =
   \frac{1}{2}\big\{1 +  {\mathop{\rm cos}\nolimits} (\frac{{{\omega _p}}}{2}T)\exp[{{ - \frac{{(\sigma _ + ^2 + \sigma _ - ^2)}{T^2}}{8}}}] \\
  & - \frac{{1 }}{2}\exp( - \frac{{\sigma _ - ^2{\tau^2}}}{2})  -\frac{{1 }}{2}\exp[ - \frac{{\sigma _ - ^2{{(\tau  - T)}^2}}}{2}]  \\
  & - {\mathop{\rm cos}\nolimits} (\frac{{{\omega _p}}}{2}T)\exp( - \frac{{\sigma _ + ^2{T^2}}}{8})\exp[ - \frac{{\sigma _ - ^2{{(\tau  - T/2)}^2}}}{2}]\big\},
\end{split}
\end{equation}
where $\omega _p$ is the pump angular frequency,  $\sigma _+$ is the width of the biphoton in the $\omega_s + \omega_i$ direction (see SM Section II for details).

To estimate the precision of measuring the time delay $T$, we now evaluate the Fisher information (FI) \cite{Liu_2020}. According to the Cramér–Rao bound \cite{Cramer1999, Lyons2018SA}, the variance $\delta^2$ of any unbiased estimator is bounded by
\begin{equation}\label{Eq4}
\delta^2 \ge \frac{1}{N \mathcal{F}},
\end{equation}
where $N$ is the number of independent experimental repetitions, and $\mathcal{F}$ denotes the Fisher information \cite{DemkowiczProgressinOptics2015}. The classical Fisher information is calculated as 
\begin{equation}\label{Eq5}
\mathcal{F}(\tau ,T) = \frac{{{{[{\partial _T}P(\tau ,T)]}^2}}}{{P(\tau ,T)[1 - P(\tau ,T)]}}, 
\end{equation}
where $P(\tau ,T)$ can be $P_\text{SHOM}(\tau ,T)$  or  $P_\text{HOM}(\tau ,T)$.

By fixing the time delay $\tau$ and scanning $T$ (e.g., by heating the sample to increase the transmission time $T$), the HOM interference pattern exhibits a nearly flat response, as shown in Fig. \ref{Fig:concept}(d). As a result, the corresponding FI remains low, on the order of 30 $\rm {ps}^{-2}$, as indicated in Fig. \ref{Fig:concept}(f).
In contrast, the SHOM interference exhibits  fluctuations in coincidence probability, ranging from 0 to 1, as shown in Fig. \ref{Fig:concept}(e). This leads to a significantly enhanced FI of approximately 5 $\times 10^6$ $\rm ps^{-2}$, as depicted in Fig. \ref{Fig:concept}(g). This represents an improvement of five orders of magnitude over the standard HOM configuration.
This dramatic increase in Fisher information arises from the additional interference term introduced in Eq. (\ref{Eq:SHOM}).
Such an enhancement provides a substantial advantage for arm length difference  measurements. With a much higher FI, the same estimation precision can be achieved with orders of magnitude fewer repetitions, thereby enabling rapid characterization of samples.

We now experimentally compare standard HOM interference with SHOM interference. The experimental setup is depicted in Fig. \ref{Fig:concept}(a), and a more detailed description of our experiment can be found in SM Section III. As comparison, HOM interference is first performed without any sample inserted. The corresponding interference pattern is shown in gray in Fig. \ref{Fig:result}(a). The measured dip exhibits a full width at half maximum (FWHM) of 0.87(2) ps, with its center position calibrated to zero delay. Subsequently, a glass sample 
is fully inserted into the signal arm, such that all signal photons pass through the sample. The resulting HOM interference pattern is shown in red in Fig. \ref{Fig:result}(a). By fitting the interference patterns, we extract the center positions of the two dips. A shift from 0 to  3.868(7) ps is observed after the sample insertion. 

Next, we present the results of SHOM interference by partially inserting the glass sample into the signal arm. A thermoelectric heater based on the Peltier effect is attached to the sample to control its temperature. 
As the temperature increases, the the optical path of the sample increase  linear with respect to temperature \cite{note1}. 
Figure \ref{Fig:result}(b) displays SHOM interference patterns measured at various temperatures. As the temperature rises from \qty{28}{\degreeCelsius} to \qty{52}{\degreeCelsius}, the central feature of the interference pattern evolves from a bump into a dip. These interference features allow us to estimate the effective optical path change of the sample, induced by the temperature change. Using the procedure described below and also in the SM Section IV to VII, we can extracted the sample optical path change at different temperatures from the measured interference patterns.

\begin{figure}[!t]
\includegraphics[width=0.48\textwidth]{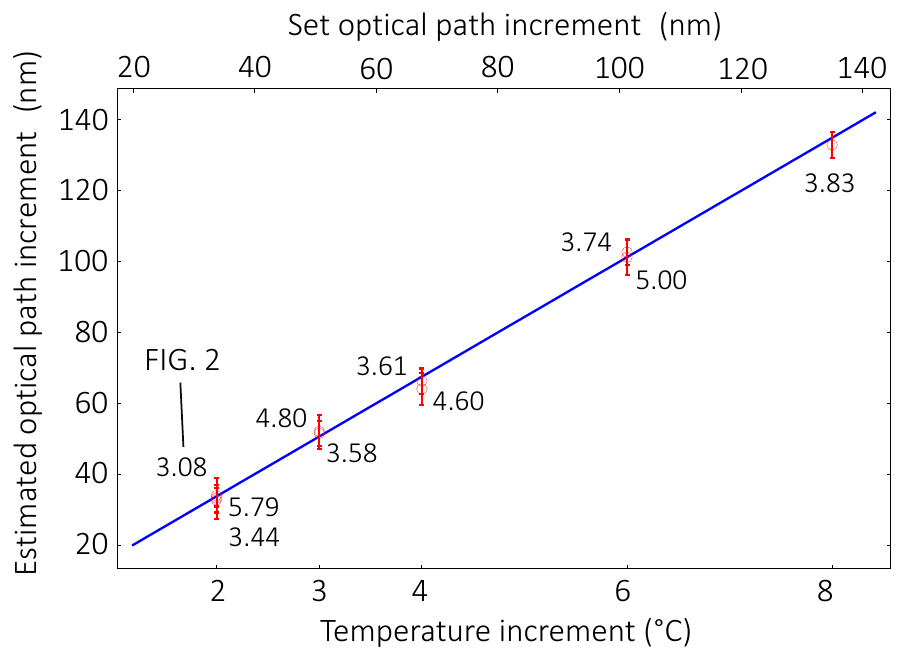}
\caption{\textbf{Estimated optical path increment versus temperature increment.} The bottom axis represents the sample temperature increment, which corresponds to set optical path increment indicated on the top axis. The red hollow circle denotes the average of the measurement results. The precision (standard deviation) for each set is indicated in the figure. }
\label{Fig:S-E result}
\end{figure}

Specifically, we first fixed the time delay $\tau$ at the central interference position and varied the sample temperature from \qty{22}{\degreeCelsius} to \qty{62}{\degreeCelsius}, effectively scanning the photon transmission time $T$. The measured results are shown in Fig. \ref{Fig:result}(c), which exhibits a sinusoidal pattern in agreement with theoretical predictions. 
According to Eq. (\ref{Eq:SHOM}), when $\tau = T/2$, the interference probability $P_{\rm SHOM}$ is primarily governed by the term $\cos\left(\omega_pT/2 \right)=\cos\left(\pi c T/\lambda_p \right)$, which exhibits  a period of $2\lambda_p$. $\lambda_p = 405.0(1)\,\mathrm{nm}$ denotes the center wavelength of the pump laser, while  $cT$ represents the optical path length, where  $c$ is the speed of light.
The peak-to-valley distance in Fig. \ref{Fig:result}(c) therefore corresponds to an optical-path change of $\lambda_p$.
Furthermore, by fitting the interference patterns in Fig. \ref{Fig:result}(c) with temperature as the horizontal axis, we obtain a period of \qty{24.03(2)}{\degreeCelsius}. This indicates that an optical path increment of 405.0(1) nm occurs over this temperature range.
Consequently, an increase of \qty{2}{\degreeCelsius} corresponds to an optical-path increase of 33.7(3) nm, which we use as the ``true”  value for later comparison.
Figure \ref{Fig:result}(f) shows the Fisher information calculated from the interference data in Fig. \ref{Fig:result}(c). We note that the Fisher information measured in Fig. \ref{Fig:result}(f) are different from Fig. \ref{Fig:concept}(g), due to the practical experimental conditions, such as imperfect visibilities (see also SM Section VIII).  Nevertheless, the maximum Fisher information from our experiment is found to be $3.40 (6) \times 10^6\,\mathrm{ps^{-2}}$, indicating high parameter sensitivity in this region. To investigate this further, we focus on measurements at \qty{38}{\degreeCelsius} and \qty{40}{\degreeCelsius}, where the Fisher information is relatively high. 

Figure \ref{Fig:result}(d) shows coincidence counts collected at these two temperatures over 200 repeated trials. From these data, we can now estimate the transmission time difference $\Delta T = |T_1 - T_2|$, which in turn yields the corresponding 
optical path change of  $ c\times  \Delta T=33.89 \pm 3.06$ nm,
as shown in Fig. \ref{Fig:result}(e). 
We now compare this newly estimated optical path change with the known true value. By subtracting the estimated value from the known set value, we can determine the measurement accuracy. 
After repeating this procedure ten times, we achieve
an average accuracy of  1.22 nm (4.07 as)  (see the SM Section XI for more details).
Note the reported attosecond resolution is inferred from delay measurements in the SHOM interferometer, not from direct photon timing. 
We also analyzed the stability of Fig. \ref{Fig:result}(e) using the tool of Allan deviation \cite{Allan1966, riley2008handbook,guo2024time}, and the averaged Allan deviation from 20 s to 70 s is calculated to be 0.87 nm (see SM Section IX for details).

Finally, the experimental results are summarized in Fig. \ref{Fig:S-E result}. The horizontal axis represents the sample temperature increment, which corresponds to increment in optical path (the ``true'' value), while the vertical axis shows the optical path increment estimated from the coincidence-count measurements. The close agreement between the two sets of values validates the accuracy and precision of our SHOM-based metrology technique. We also present a comparison between SHOM and HOM interference in Table \ref{Tab:compare}. The SHOM interference exhibits improved accuracy and precision, while requiring significantly fewer photons—reducing the photon numbers from    $O(10^{11})$ to $O(10^7)$  and achieving this with 175 times fewer experimental trials.

\begin{table}[!t]
\caption{Comparison of HOM and SHOM interference. AA is the average accuracy and AP is the average precision.} 
\centering 
\begingroup
\renewcommand{\arraystretch}{1.4} 
\setlength{\tabcolsep}{0pt} %
\begin{tabular}{p{0.08\textwidth} p{0.08\textwidth} p{0.08\textwidth} p{0.08\textwidth} p{0.08\textwidth} p{0.08\textwidth}}
\hline
\rowcolor{lightyellowgreen}
\toprule
{} & \textbf{\makecell{Set up}} & \textbf{\makecell{Trials}}& \textbf{\makecell{Photons}} & \textbf{\makecell{AA\\(nm)}} & \textbf{\makecell{AP\\(nm)}} \\
\midrule
\hline

\makecell{Ref. \cite{Lyons2018SA}} & \makecell{HOM} & \makecell{35000} & \makecell{\raisebox{0.3ex}{$O(10^{11})$}} & \makecell{1.7} & \makecell{4.8 }  \\ 

\makecell{This work} & \makecell{SHOM} & \makecell{200} & \makecell{$O(10^{7})$} & \makecell{1.22 } & \makecell{4.09}  \\
\hline
\bottomrule
\end{tabular}
\endgroup
\label{Tab:compare}
\end{table}

In SHOM interference, the two dips interfere to produce a rapidly varying structure—either a dip or a bump—at the center of the interferogram \cite{Strekalov1998PRA, Pittman1996PRL}. This central feature underlies the quantum enhancement observed in the reported measurement. The characteristic dip–bump–dip structure has been widely reported in quantum optical coherence tomography (QOCT) studies \cite{Strekalov1998PRA, Nasr2003PRL, Abouraddy2002PRA, Li2023OLT}. In such contexts, the central bump is often labeled an “artifact” \cite{lavoieQuantumopticalCoherenceTomography2009}, as it complicates quantum imaging by potentially obscuring the true position of the reflection interface, especially when the sample structure is not known  in advance. Several strategies have been developed to suppress this artifact, including pump frequency dithering \cite{Abouraddy2002PRA, nasr2004dispersion}, the use of broadband pump lasers \cite{gracianoInterferenceEffectsQuantumoptical2019}, and machine learning approaches \cite{kolenderska2021artefact, maliszewski2021artefact}. In contrast, our work turns this typically undesirable feature into a resource for high-precision optical path metrology.
We emphasize that, although our demonstration focuses on measuring the sample's optical path difference rather than its thickness, the proposed strategy—namely, the sample-half-inserted quantum interferometer—still offers advantages over the traditional sample-full-inserted approach when applied to the same sample. To further estimate the thickness of the sample, it would require more systematic characterization of the sample properties, e.g., the thermal expansion coefficient and the thermal-optic coefficient.

Regarding the ultimate achievable precision, the Cramér–Rao bound dictates that the measurement precision scales with the inverse square root of the number of repetitions (see SM Section X for detailed discussion). Thus, higher precision can in principle be achieved by increasing the number of measurements, provided the noise remains white. The ultimate limit, however, is set not by the SHOM method itself but by instrumental factors—such as the resolution and repeatability of the stepper motor, the stability of the temperature control, and the timing jitter of the APDs and TIA. Further improvements in precision are therefore possible with better instrumentation and more measurements. Further more, we discuss the robustness of the SHOM interferometer to variations in photon splitting ratio in SM Section XIII.

In conclusion, we have proposed and demonstrated an elegant way for optical path difference measurement using SHOM interference. The Fisher information obtained from a single SHOM trial exceeds that of previous HOM-based methods by five orders of magnitude, enabling a two-order-of-magnitude improvement in measurement precision. 
Experimentally, we achieved nanoscale precision and accuracy with only $O(10^7)$ photon numbers.
In the future, the ``sample-half-inserted” operation could be adapted as a versatile strategy for metrology in other quantum interferometers, such as N00N-state or Franson interferometers. 
Notably, measurements of changes in the relative optical path length constitute a promising approach to practical, quantum-enhanced metrology of transparent materials. Potential applications include quantum optical coherence tomography \cite{Abouraddy2002PRA, Nasr2003PRL}, surface profiling for microscale manufacturing \cite{Whitehouse2023handbook}, and thickness monitoring of thin-film deposition in optical coating and semiconductor fabrication \cite{Piegari2018book}.

Note: After the submission of this work, we became aware of a related study by Lualdi \textit{et al.}~\cite{lualdi2025ScienceAdvances}, which reports attosecond-level precision using highly nondegenerate energy-entangled states, in contrast to the sample-half-inserted approach employed in our work.

\section*{Acknowledgments}
This work was supported by the National Natural Science Foundations of China (Grant Nos. 12421005,  12574389, 92365106, and 92365107),  the National Key R$\&$D Program of China (Nos. 2024YFE0102400, 2019YFA0308700, and 2019YFA0308704),  the Hunan Major Sci-Tech Program (Grant No. 2023ZJ1010), and the Program for Innovative Talents and Teams in Jiangsu (Grant Number JSSCTD202138). C. Y. and O. S. M. L. acknowledge support from the National Science Foundation (NSF) under Award No. 2517031.
We thank Prof. Yan Guo, Prof. Baihong Li, and  Prof. Ruifang Dong for helpful discussions.

\bibliographystyle{apsrev4-2}

\end{document}